\documentclass[conference]{IEEEtran}

\ifCLASSINFOpdf
\else
\fi

\usepackage[bookmarks=false]{hyperref}
\usepackage{graphicx}
\usepackage{color}
\usepackage{subfig}
\usepackage{subfig}
\usepackage{morefloats}
\usepackage{caption}

\begin{document}

\title{Complementary Training Programme for\\Electrical and Computer Engineering Students Through an Industrial-Academic Collaboration\\(Extended Version)}

\author{\IEEEauthorblockN{Felipe R. Monteiro, Phillipe A. Pereira, Lucas C. Cordeiro, Cicero F. F. Costa Filho and Marly G. F. Costa}
\IEEEauthorblockA{Electronic and Information Research Centre\\
Federal University of Amazonas -- Manaus, Amazonas, Brazil\\
\{rms.felipe, apphillipe, lucasccordeiro, cffcfilho, marlygfcosta\}@gmail.com}}

\maketitle

\begin{abstract}
We describe the results of an industrial-academic collaboration among the Graduate Program in Electrical Engineering (PPGEE), the Electronics and Information Research Centre (CETELI), and Samsung Eletr\^onica da Amaz\^onia Ltda. (Samsung), which aims at training human resources for Samsung's research and development (R\&D) areas. Inspired by co-operative education systems, this collaboration offers an academic experience by means of a complementary training programme (CTP), in order to train undergraduates and graduate students in electrical and computer engineering, with especial emphasis on digital television (TV), industrial automation, and mobile devices technologies. In particular, this cooperation has provided scholarships for students and financial support for professors and coordinators in addition to the construction of a new building with new laboratories, classrooms, and staff rooms, to assist all research and development activities. Additionally, the cooperation outcomes led to applications developed for Samsung's mobile devices, digital TV, and production processes, an increase of $37\%$ in CETELI's scientific production ({\it i.e.}, conference and journal papers) as well as professional training for undergraduates and graduate students.\\
\end{abstract}

\begin{IEEEkeywords} co-operative education; complementary training programme; engineering students; extra-curricular programmes; university and industry collaboration. \end{IEEEkeywords}

\IEEEpeerreviewmaketitle

\section{Introduction}
\label{sec:introduction}

Due to the fast technological progress in the past decade, companies are facing a degree of technological complexity that requires more specialised engineers to push forward in their research and development (R\&D) projects~\cite{Maraghy2011}. For this particular reason, foremost universities around the globe tend to constantly revise the education of science, technology, engineering and mathematics (STEM) fields, in order to be up-to-date with industry's requirements and adequately preparer their engineering students for their future jobs~\cite{cbi2014, vanderHoek2006}. However, in Brazil, this might not be a simple task. In particular, the curricula of Electrical and Computer Engineering courses from Brazilian universities follow restricted requirements established by the Brazilian Ministry of Education, which complicates curricula modifications and makes its progression an arduous task~\cite{lucena:2011}. In order to tackle such problem, universities pursue for partnerships with industry in order to invest in co-operative education programmes to fulfil the gap in the educational process of their engineering students~\cite{lucena:2011, ref:upgrading2011, waldir:2015}. In addition, many studies have shown the importance of work experience during college for the professional development of engineering students and its positive impacts on local economy~\cite{powell:2001, Schuurman:2005}. 

In this scenario, as an alternative to a close convergence between industry and academia in addition to promoting innovative R\&D projects to benefit the engineering students, Federal University of Amazonas (UFAM) founded the Electronic and Information Research Centre (CETELI). It has $11$ years of experience in collaborating with several companies located at the Industrial Pole of Manaus (PIM). Indeed, its mission is to promote research, technological progress, and human resource training in Amaz\^onia with the purpose of achieving excellence in the fields of electronic and information technology, industrial automation and biomedical engineering~\cite{costa:2011}. Thus, PIM's companies have funded several R\&D projects in such fields. Some companies that collaborated with CETELI include Tr\'opico Systems and Telecommunications~\cite{tropico}, for developing telecommunication systems; Nokia Institute of Technology (INdT)~\cite{indt}, for developing mobile applications and software verifiers; and currently Samsung~\cite{samsung}, for training students and developing applications to mobile devices, digital television (TV), and industrial automation~\cite{ref:upgrading2011}.

The projects developed at CETELI are typically coordinated by (permanent) professors from UFAM, with the mission to achieve goals defined by each specific cooperation. Moreover, they aim to train human resources in undergraduate and graduate levels in different engineering domains. In particular, CETELI has already prepared students who were able to start working in the industry after concluding their courses as well as researchers in technological innovation who were able to contribute in the development of innovative products for the market~\cite{ref:upgrading2011, costa:2011, waldir:2015}. CETELI has also a number of products available to end users and customers, {\it e.g.}, hardware and software products developed by engineering students and professionals for Tr\'opico Systems, INdT, and Samsung~\cite{ref:cicero2010}. The development of these projects contribute to the high quality of the academic education at UFAM and it also builds a reliable partnership with companies.

This paper addresses three major contributions: 

\begin{itemize}
\item First, we report the effort to establish an industrial-academic cooperation between CETELI, UFAM and Samsung, in order to meet the demand for trained human resources according to the (current) market interests and company needs;

\item Second, we present the proposed co-operative education programme, called as Complementary Training Programme (CTP) and its educational structure from the theoretical and practical perspective;

\item Finally, we highlight the CTP's accomplishments and their impacts in the educational process of the undergraduate and graduate engineering students.
\end{itemize}

As aforementioned, such an industrial-academic collaboration is focused on research areas related to mobile devices, digital TV, and industrial automation, due to industry demand. For instance, mobile device technologies ({\it e.g.}, smartphones and tablets) grow every year worldwide by improving key-features such as processing power and storage capacity, which results in a better support to a wide range of applications ({\it e.g.}, games, social interaction and services that were previously restricted to computers only)~\cite{ref:lucas2008, ref:lucas20082}. Additionally, digital TV has shaped a new application market for user's interactivity, especially in Brazil, where such technology has been recently implemented~\cite{prata:2015}, which brings a new range of research and application development challenges. Also, industrial automation is an emerging area, due to concepts such as the interconnected Hybrid Internet of Things (IoT)~\cite{ref:iot2011, ref:Vermesan2014}, allowing real-time communication, which thus increases the need for new and innovative solutions. These technological innovations have created the need for training professionals to develop such powerful applications, which demands development centres to train people in these research field.

\section{Industrial-Academic Collaboration}
\label{sec:program_organization}

The partnership agreement between CETELI, UFAM, and Samsung was established to allow the investigation of new research areas and to provide an extensive training programme, for graduate and undergraduate engineering students, based on topics related to current industry demand. However, we have faced three major challenges to accomplish this collaboration project:

\begin{enumerate}
  \item The identification of the primary goals of each partner and the establishment of a common ground in the partnership, which is of paramount importance for the success of such collaboration~\cite{pronk:2015};
  \item Overcome the bureaucracy in the Federal Institution of Higher Education (IFES), in order to establish the partnership with a private company. In fact, the respective project must be approved by the original department, the Innovation Technology Department, Directors Board, the Administrative Dean, and then the Legal Department. As one can see, it is a long journey, with no possibility of acceleration among those departments;
  \item Establish contracts on intellectual property (IP) ownership and confidentiality, which are of paramount importance for the company representatives; note that this process must be as clear as possible for all involved parties~\cite{pronk:2015}.
\end{enumerate}

In order to overcome issue {\it (i)}, the project proposal has taken into account the university's interests to build distinct knowledge, as well as Samsung's needs regarding the availability of human resources in its R\&D department, in order to design, develop, and test (innovative) products. As a result, the chosen R\&D fields included software development activities related to mobile devices, digital TV, and industrial automation. 

Regarding issue {\it (ii)}, the excessive bureaucracy of the project's approval process was attenuated by a continuous follow-up of its legal procedure at UFAM. 

Finally, with respect to issue {\it (iii)}, the official project proposal has established a three-years cooperation plan; in particular, the proposal covers the courses' contents, practical activities, scholarships, teaching instruction, project coordination, investments in infrastructure and equipment, and all legal contracts regarding IP ownership and confidentiality. In addition, UFAM accepted the minor percentual portion of the joint ownership of IP, due to the fact it is the major beneficiary of this project, once it receives the majority of incomes and infrastructure. 

\section{Complementary Training Programme}
\label{sec:training}

Inspired by co-operative systems of education~\cite{coop2008}, the ultimate goal of this industrial-academic collaboration is to implement a continuing-education model named as Complementary Training Programme (CTP). Similar to co-operative education programmes, CTP aims to combine a classroom-based learning approach with a more realistic work-based practical experience, in order to train undergraduate students in three major research areas:

\begin{itemize}
 \item Mobile devices applications development.
 \item Digital TV applications development.
 \item Emerging technologies for industrial automation systems.
\end{itemize}

It is worth noting that differently from the lecture/laboratory approach, which is usually applied to the engineering courses, this training programme delivers a work experience, where the undergraduate students work on a project from scratch until the accomplishment of a reliable product in each respective research area. In addition, another differential of the CTP is the inclusion of graduate engineering students from the Postgraduate Programme in Electrical Engineering (PPGEE) in its process. 

Through the mentorship of the undergraduate students, the graduate ones gain more teaching and leading experience, in addition to more financial incentive for their research. Indeed, Samsung provides scholarships and financial incentive for scientific publications ({\it i.e.}, all costs for conference attendance and language editing) for PPGEE students, who work on research areas related to the project's fields of interest, and it also provides financial support for the ones who work as mentors during the projects. Furthermore, Samsung financially supports professors associated with CETELI, who work as project leaders, in order to supervise graduate and undergraduate students during the development of their work.

\subsection{Programme Structure}
\label{sec:training}

\begin{figure}[h]
\centering
\includegraphics[width=0.5\textwidth]{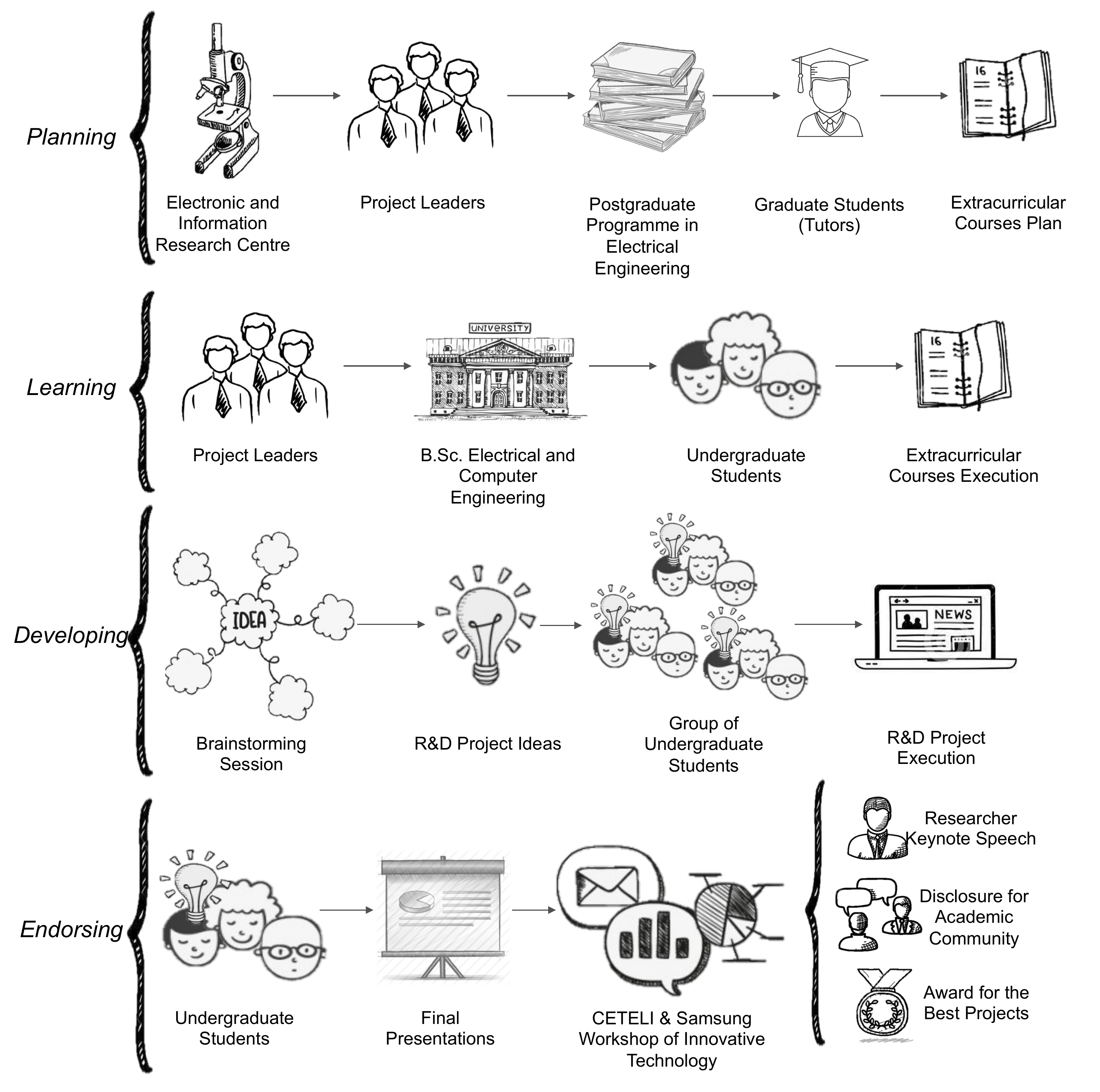}
\caption{Workflow structure of the Complementary Training Programme.}
\label{fir:ctp}
\end{figure}

CTP follows a specific workflow, which is modularised into four sub-modules identified as Planning, Learning, Developing, and Endorsing, as one can see in Fig.~\ref{fir:ctp}. Importantly, CTP's workflow is the same despite the target research area.

\noindent
{\bf Planning.} First of all, three professors associated with CETELI are elected to be project leaders in each respective research area. Then, the project leaders select on average two graduate students from PPGGE, based on their curriculum and theme of expertise, to be tutors of undergraduate students during the project. Both professors and graduate students design extracurricular courses, which must cover all key-subjects to provide the necessary background that undergraduate students need to carry out a R\&D project into the aforementioned research areas.

\noindent
{\bf Learning.} At this stage, each project leader performs an admission process to select $20$ undergraduate students, based on their grade point average (GPA) and availability. Then, during $4$ months the selected undergraduate students receive the aforementioned extracurricular courses, which are taught by the tutors (graduate students) and specialised professionals from industry. In addition, all undergraduate students are evaluated in these courses on a 10-scale grading system through courseworks and exams; indeed, they must achieve at least $7.0$ grade and $75\%$ of attendance by the end of each course, in order to continue in the protect. Importantly, each project offered from $4$ to $5$ extracurricular courses, lasting about $20$ hours each, which implies $80$ to $100$ hours of extracurricular education for undergraduate students.

\noindent
{\bf Developing.} By the end of the learning activities, all professionals and students perform a brainstorming session, in order to suggest R\&D projects ideas. Most importantly, all ideias are proposed by undergraduate students and are evaluated by the project leaders, tutors, and industry professionals according to three main aspects: originality, innovation, and feasibility. On average, $4$ project ideas are selected, so, the undergraduate students are split into groups and each group works to implement one of them. During the practical phase, graduate students and industry professionals continuously supervise the undergraduate students' practical activities. Such process is also monitored by project leaders by means of periodic meetings, which help ensure the learning progress efficiency and the quality of student's work. This stage lasts approximately $4$ months.

\noindent
{\bf Endorsing.} Once the proposed R\&D projects are concluded ({\it i.e.}, each group presents a stable and tested version of the proposed product), CETELI organises a workshop for the academic community, named as CETELI \& Samsung Workshop of Innovative Technology, where the undergraduate students present their outcomes and experiences during the project. Throughout the workshop, a committee, which is composed by project leaders, professors associated with CETELI, tutors, specialised professionals, and representatives from Samsung, evaluates each R\&D project according to its outcomes and level of innovation, in order to award the best R\&D project in each research area. Indeed, the winner students are awarded with Samsung devices, such as smartphones and smart TVs. It is worth noting that such healthy competition is important to push undergraduate students to attempt more creative and challenging ideias. Moreover, as a way to inspire the academic community, renowned researchers are invited to delivery a keynote speech about emerging technologies from each research area.

The following subsections describe the content related to each area, in addition to the goals targeted to ensure the training quality for the undergraduate students.

\subsection{Mobile devices applications development}

Projects related to this area aim to develop applications for mobile devices ({\it i.e.}, smartphones and tablets), which may vary from public utilities to games, entertainments, and content searchers. Indeed, the primary goal here is to provide the necessary background to develop high-quality mobile applications for the Android platform~\cite{ android16}. During this partnership, it was offered $3$ training in this area (each with $20$ students on average). The {\it Learning} stage of this area covered the following subjects: Java Programming Languages for Mobile Devices, Software Development Methodologies for Mobile Devices, Operating Systems for Mobile Devices, and Software Verification and Testing.

\subsection{Digital TV applications development}

This research area aims to develop digital TV applications to promote more interaction between users and contents, which may vary from interactive programs such as games, news report, and entertainments to programs with a specific collection of content, such as TV listings and utilities to support daily user activities. During this partnership, it was offered $2$ trainings in this area (each with $15$ students on average). The {\it Learning} stage of this area covered the following subjects: Fundamentals on Digital TV, Java for Digital TV, Software Development using Middleware for Digital TV, C/C++ Programming, Programming for Embedded Systems, and Embedded Linux for Digital TV.

\subsection{Emerging technologies for industrial automation systems}

This research area aims to develop automatic solutions to improve industrial production environments, in order to make its processes faster, safer, and more accurate. During this partnership, it was offered $2$ training (each with $15$ students on average). The {\it Learning} stage of this area covered the following subjects: Introduction to System Automation, Introduction to Mobile Robotics, Introduction to Industrial Robotics, System Development for Plants Automation, and Software Development for Real-time Systems.

\section{Partnership Outcomes}
\label{sec:results}

After $3$ years of partnership between CETELI, UFAM and Samsung, a total of $13$ mobile applications, $5$ digital TV applications, and $4$ industrial automation applications were developed. Additionally, $8$ master degree dissertations were defended in areas related to the project's fields of interest, as well as, $25$ scientific publications in top conferences and journals. In the following sections, the outcomes of each area area described in details.

\subsection{Mobile devices applications development}
\label{subsec:results_mobile}

In mobile devices area, $60$ undergraduate students were trained. In addition, the practical activities in this field resulted in the production of $13$ mobile applications to smartphones and tablets, which were designed to support users in different daily base activities. As example of such applications, $3$ mobile applications are described below, and it is worth noting that all $3$ are patented to ensure the students royalties.

\begin{figure}[h]
\centering
\includegraphics[width=0.35\textwidth]{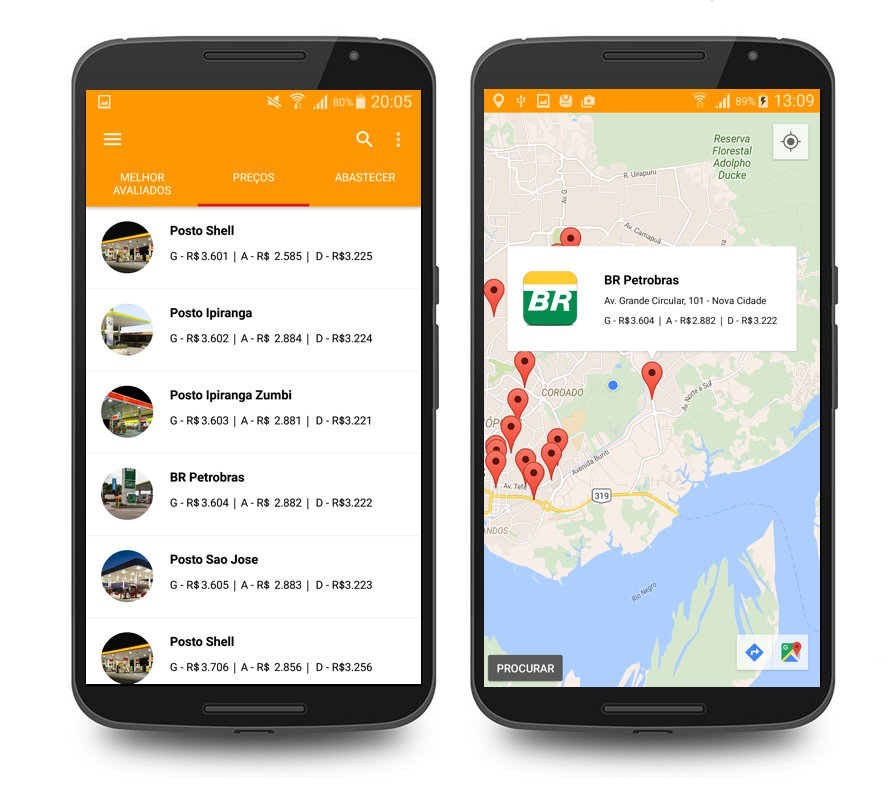}
\caption{{\bf PitStop} application's layout.}
\label{fig:app_pitstop}
\end{figure}

{\bf PitStop} application, shown in Fig.~\ref{fig:app_pitstop}, has two primary goals: to support drivers to organise all the information about their vehicles and to search for the cheapest gas station near by with the best service evaluation. The user can upload the information about his/her vehicle ({\it e.g.}, model, fuel source, data of last inspection) and, based on those information, the application can make suggestion for the driver such as if it is about time to make another inspection. In addition, the application shows the closest gas station that offers the cheapest price for the kind of fuel source that the vehicle needs. Users can search on a map for the others gas stations as well, check which services they provide, compare their prices, make evaluations about their services, among other features.

\begin{figure}[h]
\centering
\includegraphics[width=0.35\textwidth]{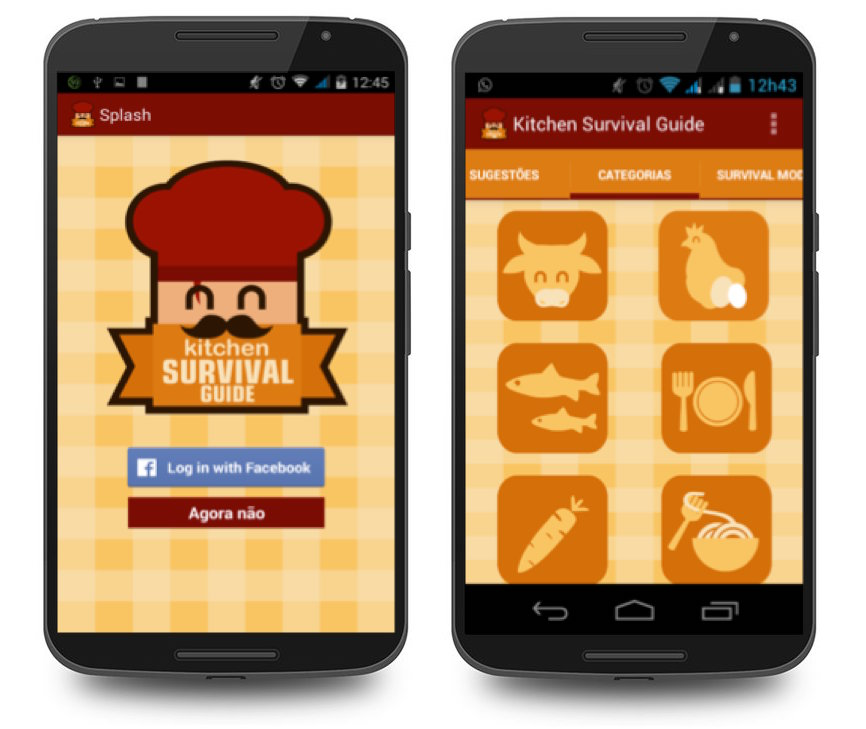}
\caption{{\bf Kitchen~Survival~Guide} application's layout.}
\label{fig:app_kitchen}
\end{figure}

{\bf Kitchen~Survival~Guide} application, shown in Fig.~\ref{fig:app_kitchen}, has the primary goal to support people who has no experience whatsoever in cooking. It differs from many others related available apps, because it provides, through a search system, recipes according to the available ingredients and household appliances that the user has at the moment. In addition, another outstanding feature of the application is its system recommendation, which automatically traces a profile for the user, detecting his/her preferences while the user uses the app, and then it recommends receipts that fit into the respective profile.

\begin{figure}[h]
\centering
\includegraphics[width=0.35\textwidth]{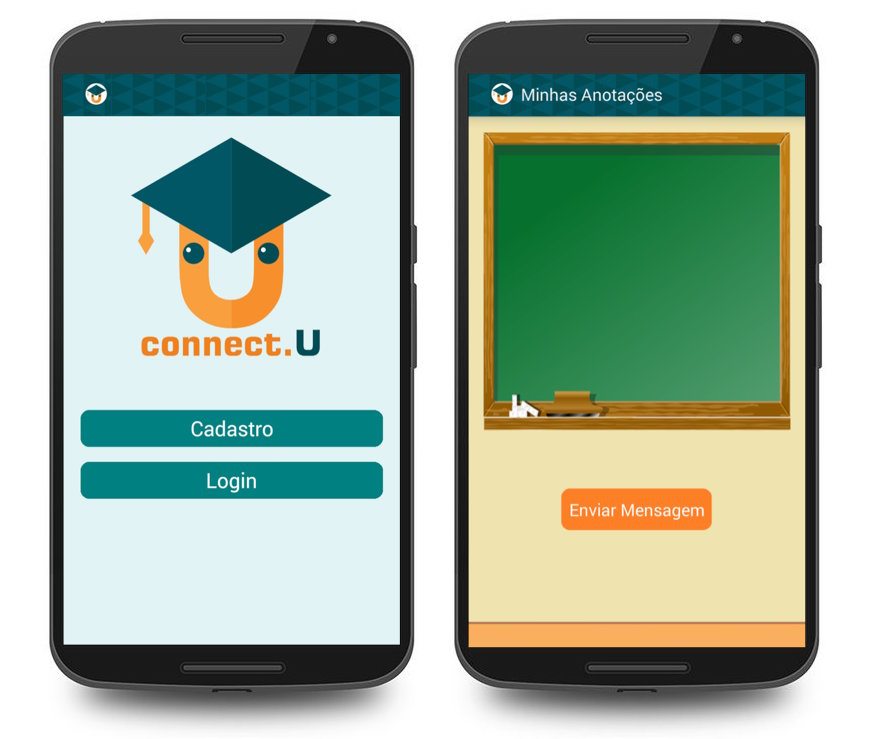}
\caption{{\bf Connect~U} application's layout.}
\label{fig:app_connect}
\end{figure}

{\bf Connect~U} application, shown in Fig.~\ref{fig:app_connect}, was developed to support students, who study at the same class and intend to share contents in a collaborative environment. It provides an area to share information, such as dates, files, exams, course-works, and all relevant material for the class progress. A place to deliberate specific topics is also available, which organises the discussions by themes. In addition, the application can be synchronised with Facebook, so the students can also publish some contents in the social media.

\subsection{Digital TV applications development}
\label{subsec:tvdigital}

In digital TV area, $28$ undergraduate students were trained. As a result of their practical activities, $5$ interactive applications for digital TV were developed. Some examples of such applications are described below, and it is worth noting that all are patented to ensure the students royalties.

\begin{figure}[h]
\centering
\includegraphics[width=0.45\textwidth]{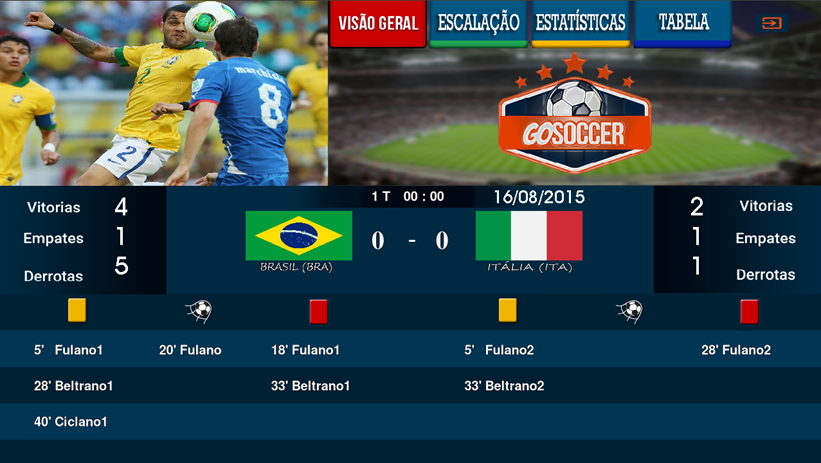}
\caption{{\bf GoSoccer} application's layout.}
\label{fig:tv_digital}
\end{figure}

Motivated by the World Cup held in Brazil in $2014$, undergraduate students have developed $2$ digital TV applications for the soccer fans. The {\bf Copa~DTV} application provided all information about the match schedules and the participants during the respective World Cup. In contrast, another developed application named as  {\bf GoSoccer}, shown in Fig.~\ref{fig:tv_digital}, provides all information about the matches from Brazilian championships in general. Such information are released during the match, so the user can track the data in real-time.

\begin{figure}[h]
\centering
\includegraphics[width=0.45\textwidth]{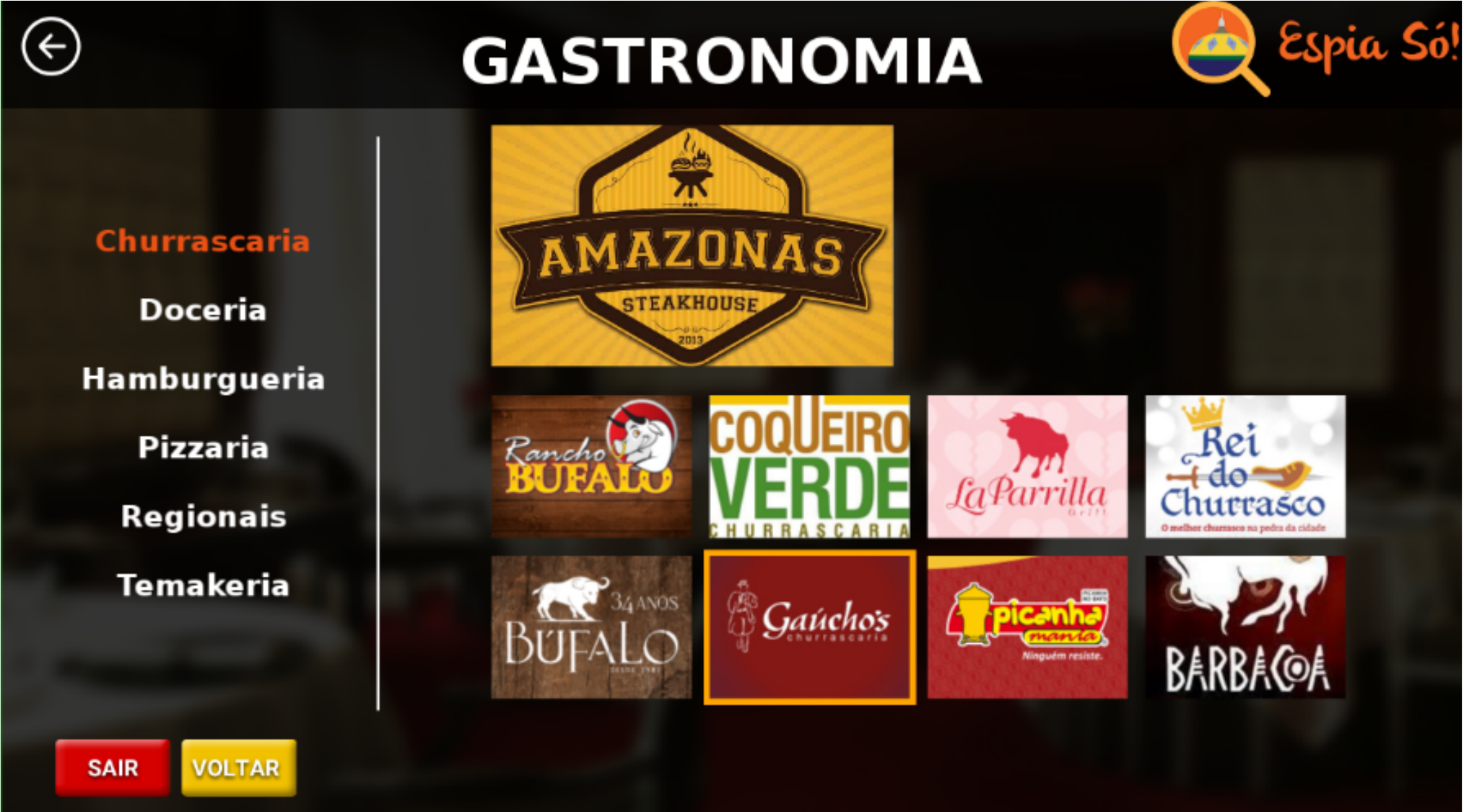}
\caption{{\bf Espia S\'o} application's layout.}
\label{fig:tv_digital1}
\end{figure}

Inspired by the touristic potential of Manaus city, undergraduate students also developed $2$ digital TV applications about what could be explored in the city. For instance, {\bf Espia S\'o} application, shown in Fig.~\ref{fig:tv_digital1}, presents an organised set of information about gastronomic and entertainment places in Manaus. It also releases information about upcoming events in the city, such as date, local, description, and so on, which are updated on a weekly or monthly basis. Another digital TV application presents information about touristic places in Manaus, providing their location and a brief explanation about each place.

\subsection{Emerging technologies for industrial automation systems}

\begin{figure}[h]
\centering
\includegraphics[width=0.4\textwidth]{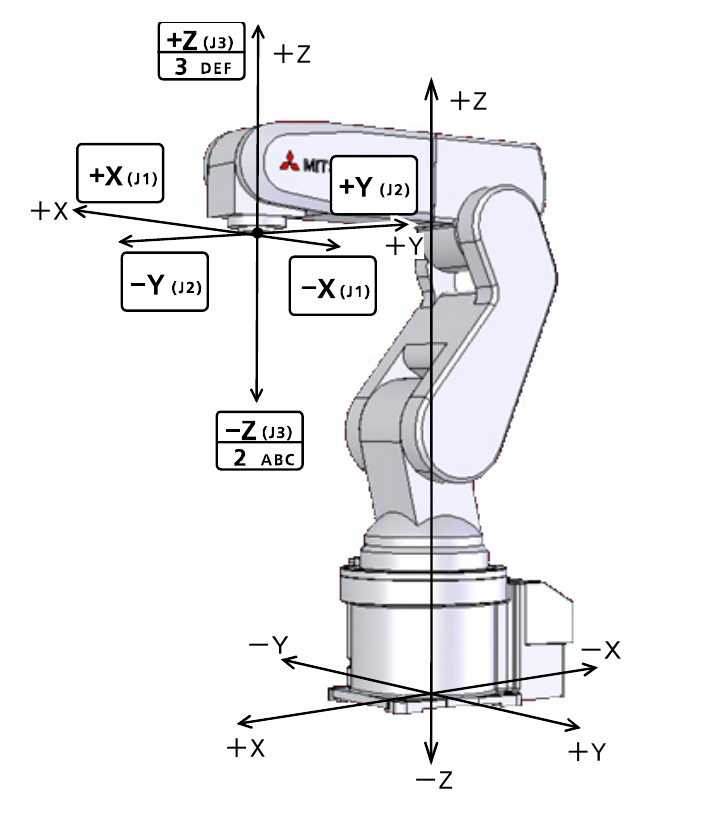}
\caption{Melfa RV-2SDB robot arm illustration and its base coordinate system.}
\label{fig:automation}
\end{figure}

In the industrial automation area, $30$ undergraduate students were trained. The students were split into $3$ classes and each one was responsible for one of the following development areas: Programmable Logic Controller (PLC) Programming, Mobile Robotics Programming, and Industrial Robotics Programming. The main goal was to apply the knowledge of each area to a global system, which comprises of a plant with two cars to transport items via five different stations. Each item enters the production line manually and the final stations are an industrial robot and a palletising station to deposit the outcomes. As an example, Fig.~\ref{fig:automation} shows the robot arm Melfa RV-2SDB~\cite{melfa2010}, which was programmed to manage the production line outcomes. Currently, this project is in its final stage of development.

\subsection{Scientific publications}

\begin{figure}[h]
\centering
\includegraphics[width=0.4\textwidth]{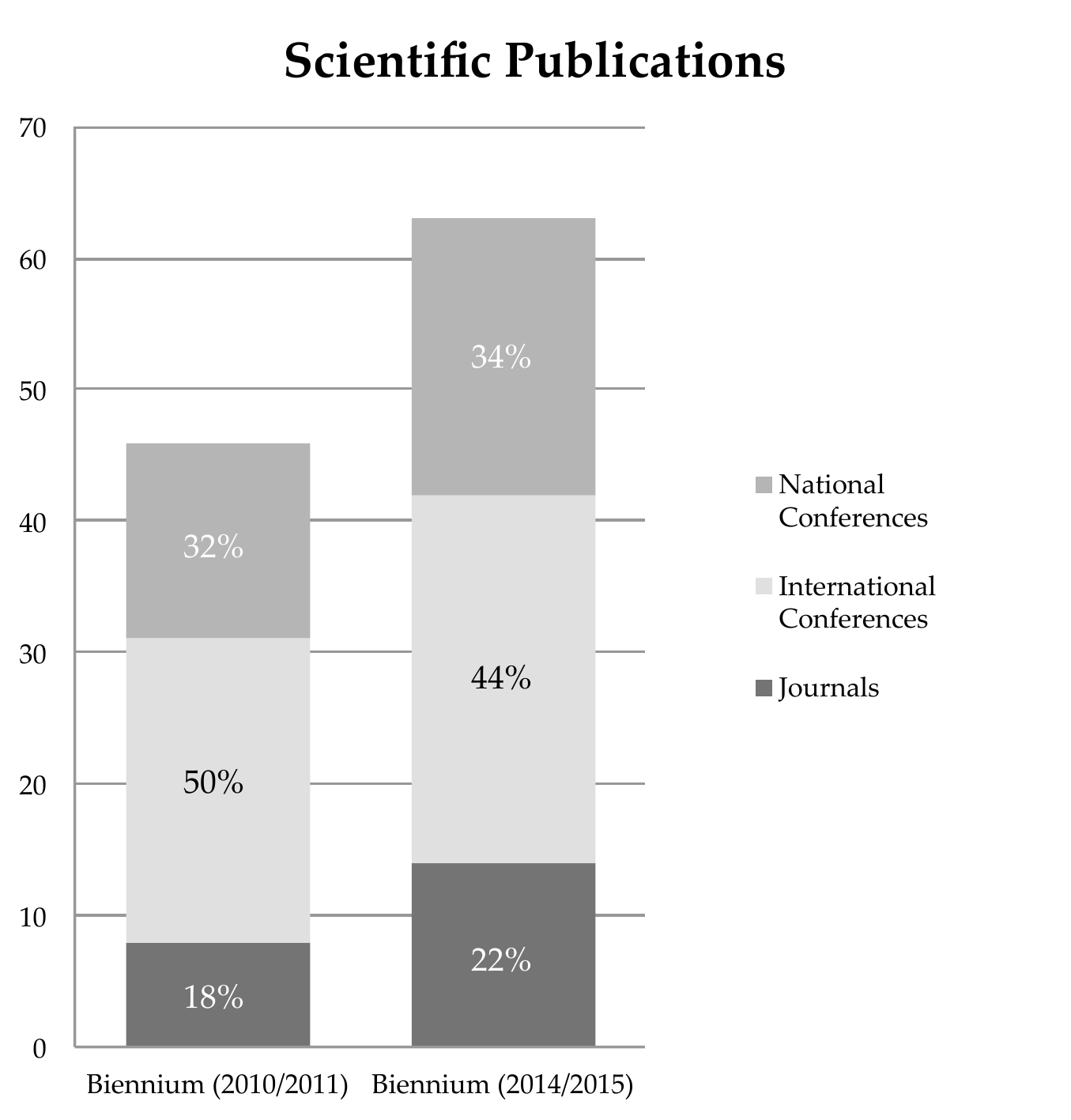}
\caption{Scientific production comparison between two periods, one without the partnership (2010/2011) and another through it (2014/2015).}
\label{fig:publications}
\end{figure}

During this partnership, an overall of $23$ articles were published in $16$ international and $7$ national conferences, in addition to $2$ journal papers~\cite{Melo:2016,Bessa:2016}; we also have other $2$ journal submissions under review. These scientific publications have contributed to specific fields related to the project's areas of interesting, such as digital TV~\cite{ref:automated2015, januario2014}, formal verification~\cite{ref:dsverifier2015, monteiro2015, mario2016, Trindade16}, and education of engineering students~\cite{waldir:2015}. As one can see in Fig.~\ref{fig:publications}, through this partnership it was possible to improve in $37\%$ CETELI's scientific production, which also implied in an increase of $29\%$ in conference participation and $43\%$ more journal publications. 

Furthermore, as a part of the program goals, four workshops were promoted to present the R\&D projects conducted by the undergraduate students, their outcomes, and to allow an open discussion with the educational community about the research topics developed by each project. Thereabout $400$ people among graduate and undergraduate students, professors, researchers, industry professionals and business representatives in general participated on the ``$1^{st}$, $2^{nd}$, and $3^{rd}$ CETELI \& Samsung Workshop of Innovative Technology'' and the ``$1^{st}$ Symposium on Quantitative Methods of Biomedical Digital Images and Biosensor''.

\subsection{Infrastructure}
\label{sec:infraestrutura}
  
One of the primary goals from the partnership between CETELI, UFAM and Samsung was the installation of a physical infrastructure for R\&D projects. In fact, a construction project to create an extension of CETELI was executed, which resulted in a new building at UFAM, named as CETELI II. Importantly, the new facilities are also used in the training process of graduate students in innovative technologies areas, extracurricular courses for undergraduate students, and all remaining project activities. Note that such facilities also provide the necessary equipment to conduct such activities.
  
CETELI II comprises $1,184m^{2}$ of area and is fully equipped with classrooms for the PPGEE, staff rooms, meeting rooms, stockrooms, and well-equipped laboratories with brand new technologies, where the R\&D projects and the extracurricular courses are conducted. In fact, the following laboratories were assembled in CETELI II: Industrial Automation, Mobile Device, Digital TV, and General Research laboratories.

   \begin{figure}[h]
   \centering
    \subfloat[Ground floor\label{floor1}]{%
      \includegraphics[width=0.50\textwidth]{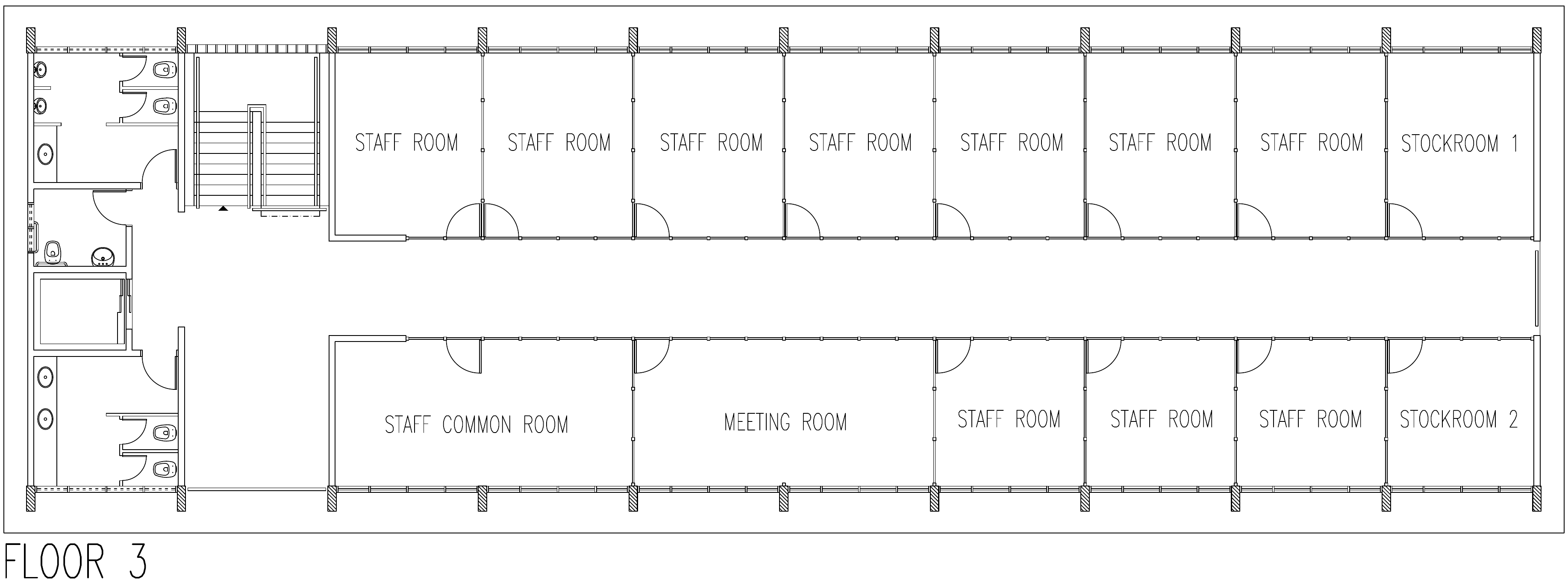}
    }
    \hfill
    \subfloat[First floor\label{floor2}]{%
      \includegraphics[width=0.50\textwidth]{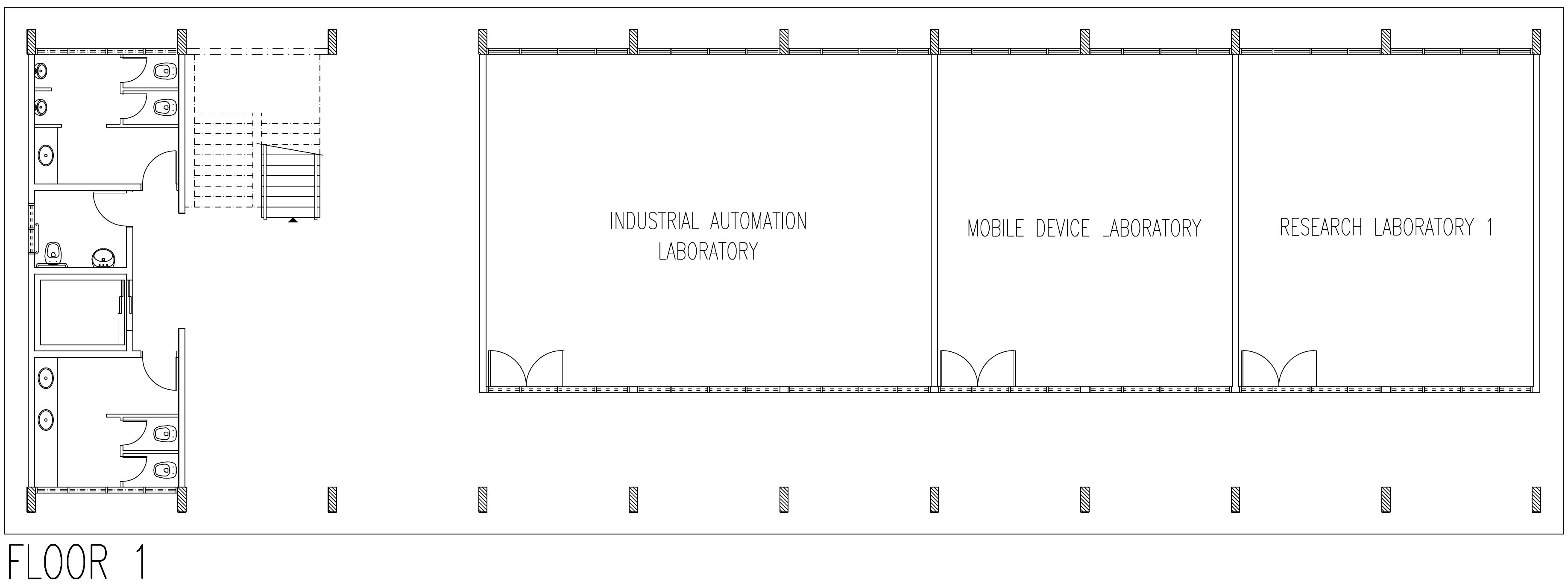}
    }
    \hfill
    \subfloat[Second floor\label{floor3}]{%
      \includegraphics[width=0.50\textwidth]{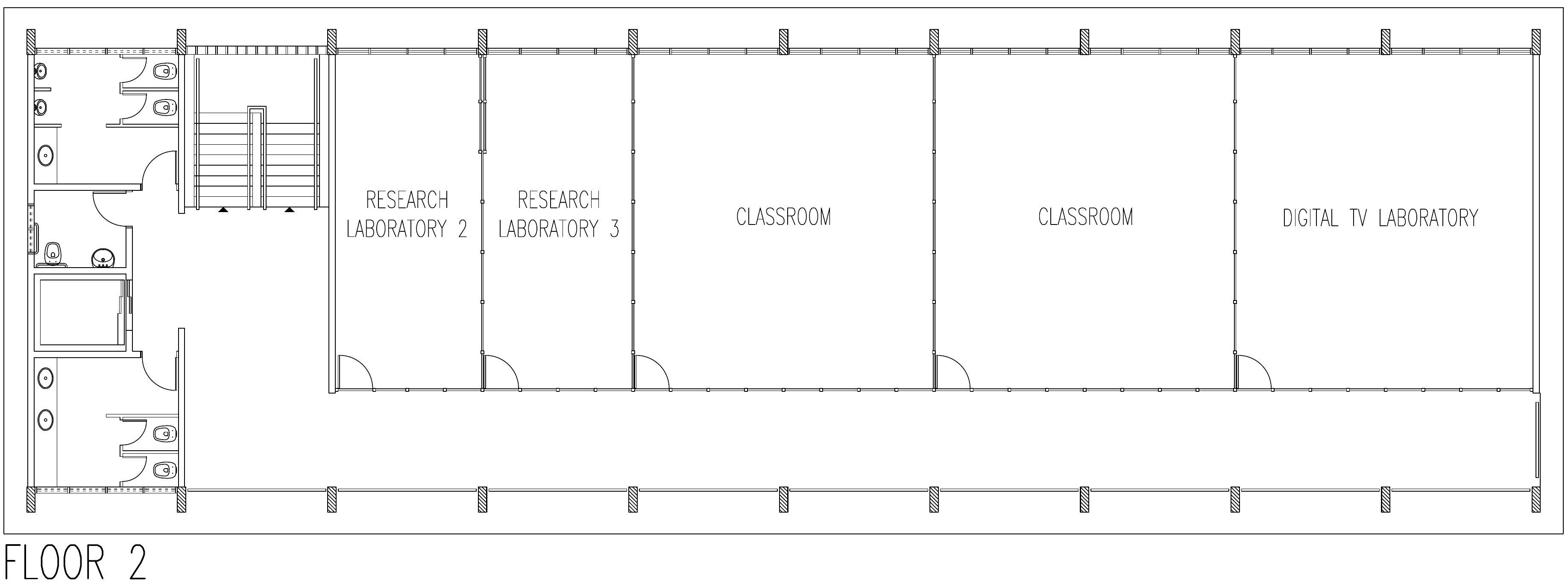}
    }
    \caption{Blueprints of CETELI II's construction project.}
    \label{fig:blueprint}
  \end{figure}

\begin{figure}[h]
\centering
\includegraphics[width=0.4\textwidth]{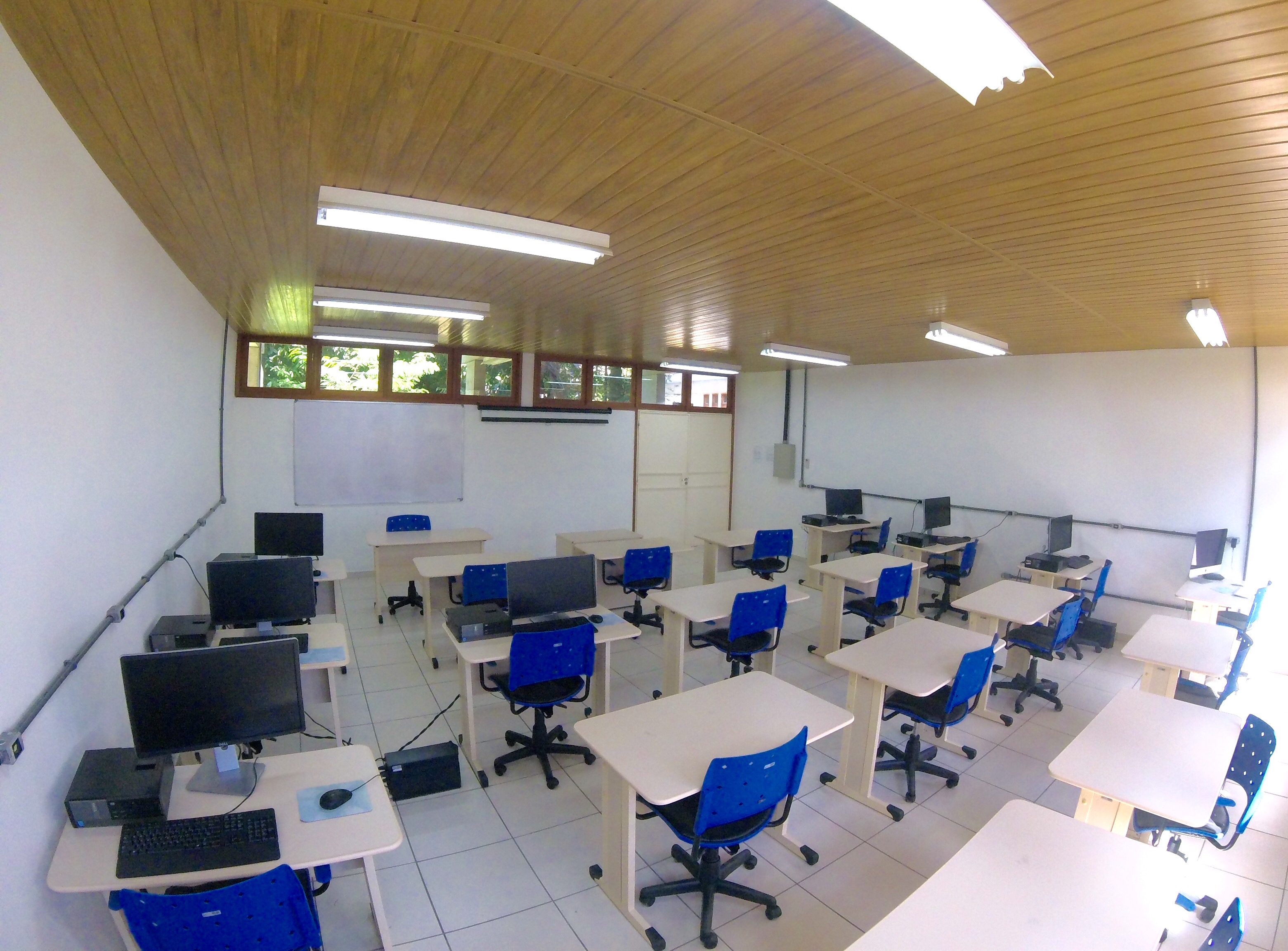}
\caption{Facilities of the Mobile Device laboratory.}
\label{fig:lab_dispositivos}
\end{figure}

CETELI II has three floors as can be seen in Fig.~\ref{fig:blueprint}, which contains the blueprints of its construction project. As one can see, Fig.~\ref{fig:blueprint}\subref{floor1} shows the blueprint of the ground floor, which contains the Mobile Device and Industrial Automation laboratories, respectively. In particular, the Mobile Device laboratory was built to provide engineering students a working environment to develop mobile applications, seeking functional and useable aspects of such systems. Its facilities, as shown in Fig.~\ref{fig:lab_dispositivos}, consist of mobile devices, $6$ standard-desktop computers, $2$ iMacs, and $2$ MacBooks (mainly used in graphic design for mobile applications). Additionally, the students have access to a wide range of different mobile-platforms ({\it e.g.}, smartphones and tablets), which were used during the development and test of their mobile applications.

\begin{figure}[h]
\centering
\includegraphics[width=0.4\textwidth]{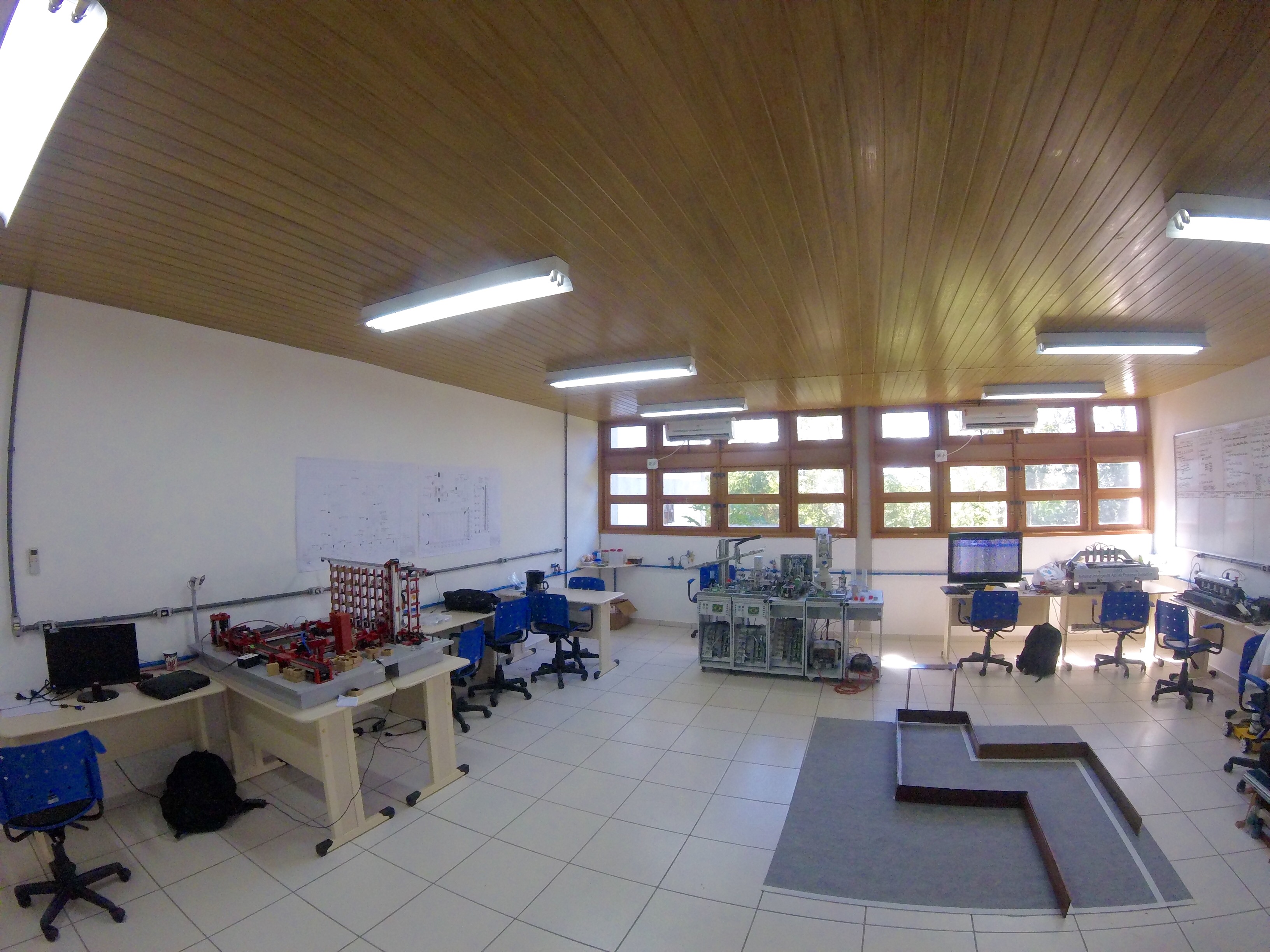}
\caption{Facilities of the Industrial Automation laboratory.}
\label{fig:lab_automacao}
\end{figure}

Industrial Automation laboratory was built to provide a working environment to research and develop automated solutions for industrial processes. As shown in Fig.~\ref{fig:lab_automacao}, its facilities consist of $6$ workstations, $4$ laptops and $4$ standard-desktop computers with real-time systems applications, in addition to robotic devices, automated platforms, and surface mount technology (SMT) component placement systems.

\begin{figure}[h]
\centering
\includegraphics[width=0.4\textwidth]{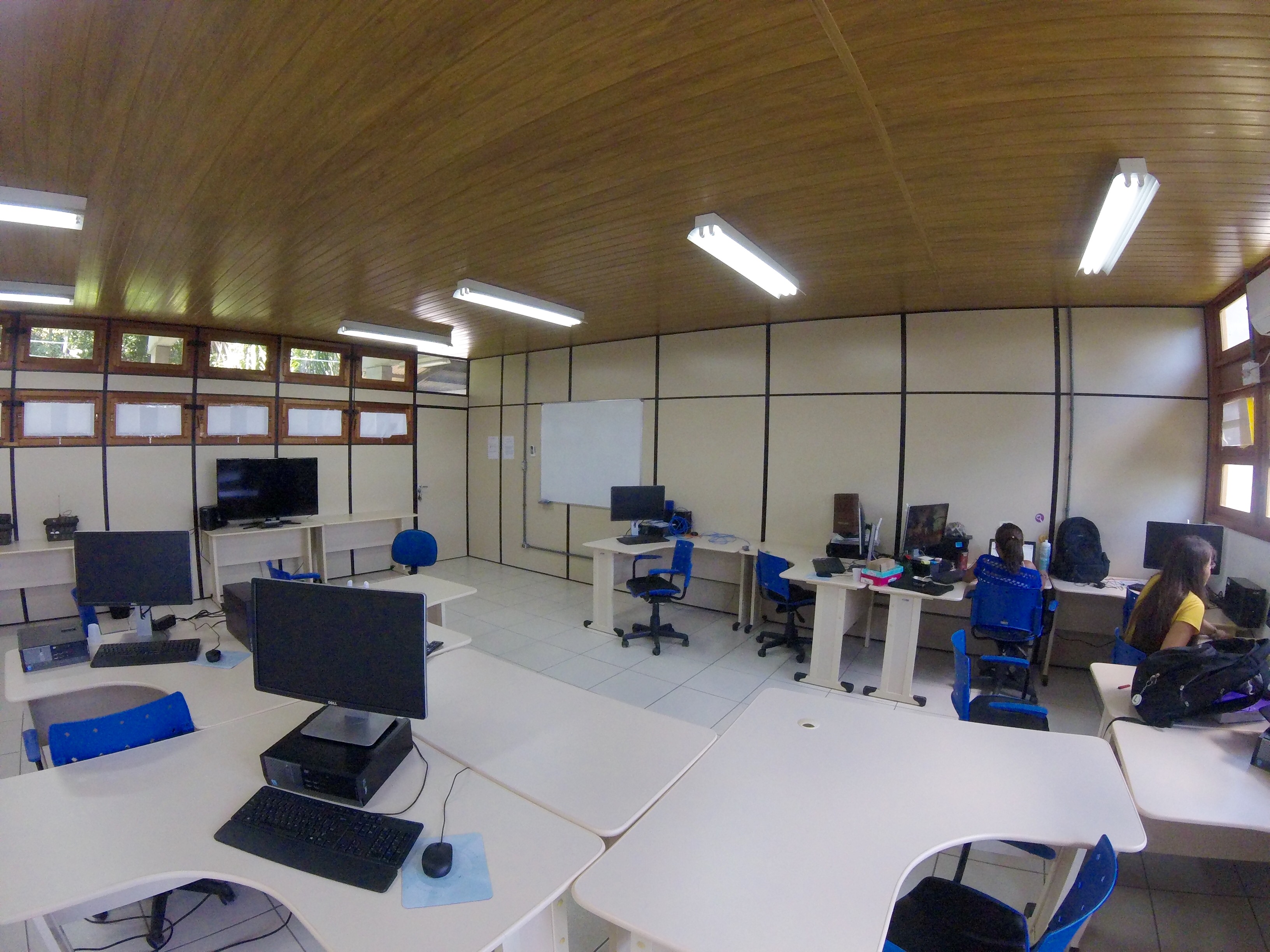}
\caption{Facilities of the Digital TV laboratory.}
\label{fig:lab_tv}
\end{figure}

Fig.~\ref{fig:blueprint}\subref{floor2} shows the blueprint of CETELI II's first floor, which comprises the Digital TV laboratory, $2$ classrooms, and $2$ research laboratories. The Digital TV laboratory was built to provide students a working environment to the development of digital TV applications, educational training and practical project activities. In addition, its facilities, as shown in Fig.~\ref{fig:lab_tv}, comprise $6$ workstations with $6$ standard-desktop computers, $3$ plasma TVs, $6$ set-top boxes, $3$ laptop computers, in addition to digital video recorders, video transmitters, and embedded software platforms.

The classrooms were built with multimedia equipment to provide interactive lectures for $20$ students. As aforementioned, these classrooms are used for educational training ({\it e.g.}, extracurricular courses), as well as, lectures for the Master of Science (M.Sc.) programme in electrical engineering from PPGEE. In addition, the remaining research laboratories were built for both undergraduate and graduate engineering students to have the opportunity to interact with each other and to tackle research challenges related to each project's field of interest.

Finally, Fig.~\ref{fig:blueprint}\subref{floor2} shows the blueprint of CETELI II's second floor, which comprises $10$ staff rooms, $2$ stockrooms, a meeting room, and a common staff room. Such facilities were built to provide professors a place to meet all students during the educational training and practical phase of each project. Furthermore, undergraduate students can use such offices for specific studies, as well as, graduate students to prepare their training material for the extracurricular courses (cf. Sec.~\ref{sec:training}). Each room contains standard-desktop computers and the necessary furniture to accommodate both professors and students. It is worth noting that CETELI II has become UFAM's patrimony and all its facilities are exceptionally available for all undergraduate and graduate students from the Faculty of Technology.

\section{Conclusions}

UFAM, CETELI and Samsung accomplished an industrial-academic collaboration with an outstanding investment to the professional training of human resources in innovative technological areas. Here, graduate and undergraduate engineering students are integrated into a Complementary Training Programme, which is inspired by co-operative educational programmes. On one hand, undergraduate students participated in training courses and had the opportunity to apply their knowledge on real R\&D projects in a professional level. It is worth noting that such approach provides to undergraduate students a learning and working experience closer to industry reality. On the other hand, graduate students were able to get involved in the aforementioned R\&D projects as well, which provided them a teaching/leading experience through the activities with the undergraduates. In addition, they received financial support to produce scientific papers, and participate in national and international conferences, which helped to increase CETELI's scientific production in $37\%$.

Such investments also allowed a wide dissemination of the R\&D projects held by CETELI and UFAM, which endorses the work quality produced in the university and its professionals. Furthermore, the infrastructure named as CETELI II represents one of the main contributions of this partnership, once its facilities and equipments will continue to be used by the undergraduate students and PPGEE's graduate students. Most importantly, the laboratories will also continue receiving investments to keep on going the research and development projects of the respective fields of interest.

Summing up on numbers, the partnership qualified $115$ undergraduate students in the CTP training in mobile devices applications development, digital TV applications development, and emerging technologies for industrial automation systems. Regarding PPGEE, $8$ graduate students obtained a Master's degree and $5$ others still have their graduate programme in progress. Additionally, $23$ conference and $2$ journal papers were published and others $2$ journal papers are currently under revision. During $3$ years of partnership, $4$ workshops were promoted in order to present the outcomes from each project and to promote discussions about emerging technologies in the aforementioned research areas. As a result, $13$ mobile and $5$ digital TV applications, in addition to $4$ industrial automation ones were developed.

These results are an example of the CETELI's potential to establish partnerships with companies, in order to improve the educational and professional experiences of its students. In particular, this partnership presented outstanding results, when compared to other investments in research and development of new technologies implemented at UFAM. From now on, with CETELI II facilities dedicated to the development of emerging technologies, research, and qualification of students, CETELI is spotted as highlight technology centre in the Industrial Pole of Manaus (Brazil).

\section*{Acknowledgment}

Part of the results presented in this paper were obtained with the project for research and human resources qualification, for under- and post-graduate levels, in the areas of industrial automation, mobile devices software, and Digital TV, sponsored by Samsung Eletr\^onica da Amaz\^onia Ltda, under the terms of Brazilian Federal Law number $8.387$/$91$.


\begin{thebibliography}{1}

\bibitem{Maraghy2011}
W.~H.~El,~Maraghy.: `` \emph{Future Trends in Engineering Education and Research}''; Advances in Sustainable Manufacturing: Proceedings of the 8th Global Conference on Sustainable Manufacturing, Springer Berlin Heidelberg, 2011, 11--16.

\bibitem{cbi2014}
Confederation of British Industry: `` \emph{Engineering our future: stepping up the urgency on STEM}''; London, England: CBI, 2014.

\bibitem{vanderHoek2006}
van der Hoek, Andr{\'e} and Kay, David G. and Richardson, Debra J.: `` \emph{Informatics: A Novel, Contextualized Approach to Software Engineering Education}''; Software Engineering Education in the Modern Age: Software Education and Training Sessions at the International Conference on Software Engineering, Springer Berlin Heidelberg, 2006, 147--165.

\bibitem{lucena:2011}
Vicente Ferreira de Lucena, Jose Pinheiro de Queiroz Neto, Joao Edgar Chaves Filho, Waldir Sabino da Silva, and Lucas Carvalho Cordeiro.: `` \emph{Gift young engineers: An extra-curricular initiative for updating computer and electrical engineering courses}''; In Proceedings of the 2011 Frontiers in Education Conference (FIE '11). IEEE Computer Society, Washington, DC, USA, 2011, S1G-1-1-S1G-6.

\bibitem{ref:upgrading2011}
C.~F.~F.~Costa Filho, O.~B.~Maia, M.~G.~F.~Costa, R.~E.~V.~Rosa, V.~L.~Lucena J\'unior, A.~M.~Gil, P.~R.~Barros, O.~S.~E.~Silva.:``\emph{Upgrading the training of undergraduate students by addressing market demands}''; Dallas, USA: Proceedings of IASTED International Conference Technology for Education, 2011.

\bibitem{waldir:2015}
W.~Sabino~da~Silva and V.~F.~de~Lucena.: ``\emph{Teaching digital TV programming for engineering students: An industry oriented proposal}''; Frontiers in Education Conference (FIE'2015), IEEE, 2015, 1--4.

\bibitem{powell:2001}
M.~Powell.:``\emph{Effective work experience: an exploratory study of strategies and lessons from the United Kingdom's engineering education sector}''; Journal of Vocational Education \& Training, Volume 53, 2001, 421--441.

\bibitem{Schuurman:2005}
M.~K.~Schuurman, R.~N.~Pangborn, R.~D.~McClintic.:``\emph{The influence of workplace experience during college on early post graduation careers of undergraduate engineering students}''; WEPAN/NAMEPA Third Joint National Conference Proceedings: Leveraging Our Best Practices: Hitting the Parity Jackpot, 2005.

\bibitem{costa:2011}
C.~F.~F.~Costa~Filho, M.~G.~F.~Costa, V.~de~Lucena~Jr., O.~S.~Melo, O.~S.~E.~Silva, and O.~B.~Maia.: ``\emph{Relatos de Casos e Experi\^encias na Educa\'c\~ao em Engenharia}''; In: Vanderli Fava de Oliveira; Carlos Almir Monteiro de Holanda; Ricardo Fialho Colares. (Org.). Engenharia em Movimento. Bras\'ilia: ABENGE, 2011, v. 1, 64--100.

\bibitem{tropico}
Tr\'opico Sistemas e Telecomunica\'c\~oes da Amaz\^onia Ltda. (2016, June). ``\emph{Tr\'opico}'' [Online]. Available: \url{http://www.tropiconet.com.br}

\bibitem{indt}
Institute of Technology Development. (2016, June). ``\emph{INdT}'' [Online]. Available: \url{http://www.indt.org.br}

\bibitem{samsung}
Samsung Eletr\^onica da Amaz\^onia Ltda. (2016, June). ``\emph{Samsung Electronics}'' [Online]. Available: \url{http://www.samsung.com/br/home/}

\bibitem{ref:cicero2010}
C.~F.~F.~Costa Filho, M.~G.~F.~Costa, V.~F.~Lucena, O.~S.~Silva, O.~Maia.:``\emph{Programa de Forma\c{c}\~ao Complementar para Alunos de Gradua\c{c}\~ao em Engenharia El\'etrica e Engenharia da Computa\c{c}\~ao}''; Fortaleza, Brazil: XXXVIII Congresso Brasileiro de Educa\c{c}\~ao em Engenharia, 2010.

\bibitem{ref:lucas2008}
L.~C.~Cordeiro, C.~Mar, E.~Valentin, F.~Cruz, D.~Patrick, R.~S.~Barreto, V.~Lucena.:``\emph{An agile development methodology applied to embedded control software under stringent hardware constraints}''; New York, USA: ACM SIGSOFT Software Engineering Notes, 2008.

\bibitem{ref:lucas20082}
L.~C.~Cordeiro, R.~S.~Barreto, M.~N.~Oliveira Jr.:``\emph{Towards a Semiformal Development Methodology for Embedded Systems}''; Funchal, Portugal: 3rd International Conference on Evaluation of Novel Approaches to Software Engineering, 2008.  

\bibitem{prata:2015}
Wilson Prata and Juan Oliveira.: ``\emph{Preferences and Concerns Regarding Mobile Digital TV in Brazil}''; 6th International Conference on Applied Human Factors and Ergonomics (AHFE 2015), Procedia Manufacturing, Volume 3, 2015, Pages 5319-5325.

\bibitem{ref:iot2011}
H.~Kopetz.:``\emph{Internet of Things}''; Boston, USA: Real-Time Systems: Design Principles for Distributed Embedded Applications, Springer US, 2011.

\bibitem{ref:Vermesan2014}
O.~Vermesan, P.~Friess.:``\emph{Internet of Things: From Research and Innovation to Market Deployment}''; Denmark: River Publishers, 2014.

\bibitem{pronk:2015}
J.~T.~Pronk, S.~Y.~Lee,	J.~Lievense, J.~Pierce, B.~Palsson, M.~Uhlen, and J.~Nielsen.: ``\emph{How to set up collaborations between academia and industrial biotech companies}''; Nature Biotechnology, Nature Publishing Group, a division of Macmillan Publishers Limited, Volume 33, (2015), 237--240.

\bibitem{coop2008}
Kettil Cedercreutz, Cheryl Cates, Anton Harfmann, Marianne Lewis, Richard Miller, Michael Zaretsky, Alexander Christoforidis, Vasso Apostolides, Anita Todd, Zach Osborne, Louis Von Eye, T. Michael Baseheart, Ann Keeling, Darnice Langford, Catherine Maltbie,  omas Newbold, Jennifer Wiswell.: ``\emph{Leveraging Cooperative Education to Guide Curricular Innovation, the Development of a Corporate Feedback Loop for Curricular Improvement}''; Cheryl Cates and Kettil Cedercreutz, University of Cincinnati, Ohio 45221, 2008.

\bibitem{android16}
Google Inc.:
\newblock {``\emph{Android Studio}''};
\newblock http://developer.android.com/sdk/index.html [accessed March-2016].

\bibitem{melfa2010}
Mitsubishi Electric Europe B.V.: ``\emph{Melfa Robots RV-2SD/2SDB Robot Arm Setup \& Maintenance}''; FA European Business Group, Germany, Tech. Rep. BFP-A8791 Version A, March 2010.

\bibitem{Melo:2016}
W.~C.~Melo, E.~B.~de Lima Filho, W.~S.~da Silva Jr.:``\emph{SEMG signal compression based on two-dimensional techniques}; Biomedical Engineering Online (Online), v. 15, p. 1, 2016.

\bibitem{Bessa:2016}
I.~V.~Bessa, H.~I.~Ismail, L.~C.~Cordeiro, J.~E.~Chaves Filho.: ``\emph{Verification of fixed-point digital controllers using direct and delta forms realizations}''; In Design Automation for Embedded Systems, v. 20, n. 2, pp. 95-126, 2016. 

\bibitem{ref:automated2015}  
F.~de~S.~Farias, W.~S.~da~Silva, E.~B.~de Lima Filho, W.~C.~Melo.: ``\emph{Automated content detection on TVs and computer monitors}''. $4^{th}$ Global Conference on Consumer Electronics (GCCE'2015), IEEE, 2015.

\bibitem{januario2014}
F. A. P. Januario, L. C. Cordeiro, E. B. de Lima Filho, V. F. Lucena Jr.: ``\emph{BMCLua: Verification of Lua Programs in Digital TV Interactive Applications}''; In $3^{rd}$ Global Conference on Consumer Electronics (GCCE'2014), IEEE, 2014, 707--708.

\bibitem{ref:dsverifier2015}
H.~I.~Ismail, I.~V.~Bessa, L.~C.~Cordeiro, J.~E.~Chaves Filho, E.~B.~de Lima Filho.: ``\emph{DSVerifier: A Bounded Model Checking Tool for Digital Systems}''; International SPIN Symposium on Model Checking of Software (SPIN'2015), 2015.

\bibitem{monteiro2015}
F.~R.~Monteiro, L.~Cordeiro, E.~de~Lima~Filho.:
\newblock {``\emph{Bounded Model Checking of C++ Programs Based on the Qt Framework}''};
\newblock In: $4^{th}$ Global Conference on Consumer Electronics (GCCE'2015), IEEE, 2015, 179--180.

\bibitem{mario2016}
M.~P.~Garcia, F.~R.~Monteiro, L.~Cordeiro, E.~de~Lima~Filho.:
\newblock {``\emph{$ESBMC^{QtOM}$: A Bounded Model Checking Tool to Verify Qt Applications}''};
\newblock In: Model Checking Software: $23^{rd}$ International Symposium (SPIN'2016), Springer International Publishing, 2016, 97--103.

\bibitem{Trindade16}
A.~B.~Trindade, and L.~C.~Cordeiro:``\emph{Applying SMT-based Verification to Hardware/Software Partitioning in Embedded Systems}''. In Design Automation for Embedded Systems, v. 20, n. 1, pp. 1-19, 2016.

\end{thebibliography}
\end{document}